\documentclass[runningheads]{llncs}
\usepackage[T1]{fontenc}
\usepackage{graphicx}
\usepackage{algorithm}
\usepackage{algorithmic}
\usepackage{amsmath}
\usepackage{amssymb}
\usepackage{subcaption}
\usepackage{svg}
\usepackage{graphicx} 
\usepackage{epstopdf}
\usepackage{float}
\usepackage{multirow} 
\usepackage{tcolorbox}
\usepackage{booktabs}
\begin{document}
\title{Uncertainty-Aware Semantic Decoding for LLM-Based Sequential Recommendation}

\titlerunning{USD for LLM-Based Sequential Recommendation}

\author{Chenke Yin\inst{1}\thanks{These authors contributed equally to this work.}\orcidID{0009-0009-2176-3967} \and
Li Fan\inst{1}\protect\footnotemark[1]\orcidID{} \and
Jia Wang\inst{1}\thanks{Corresponding author: Jia Wang. Email: \email{Jia.Wang02@xjtlu.edu.cn}} \orcidID{0000-0002-3165-7051} \and
Dongxiao Hu\inst{1}\orcidID{} \and
Haichao Zhang\inst{1}\orcidID{0000-0002-1805-1047} \and
Chong Zhang\inst{1}\orcidID{0009-0003-2020-6989} \and
Yang Xiang\inst{1}\orcidID{0009-0007-9693-7963}
}

\authorrunning{C. Yin et al.}

\institute{Xi'an Jiaotong-Liverpool University}

\maketitle

\begin{abstract}

Large language models have been widely applied to sequential recommendation tasks, yet during inference, they continue to rely on decoding strategies developed for natural language processing. This creates a mismatch between text-generation objectives and recommendation next item selection objectives. This paper addresses this limitation by proposing an Uncertainty-aware Semantic Decoding (USD) framework that combines logit-based clustering with adaptive scoring to improve next-item predictions. Our approach clusters items with similar logit vectors into semantic equivalence groups, then redistributes probability mass within these clusters and computes entropy across them to control item scoring and sampling temperature during recommendation inference. 
Experiments on Amazon Product datasets (six domains) gains of 18.5\% in HR@3, 11.9\% in NDCG@3, and 10.8\% in MRR@3 compared to state-of-the-art baselines. 
Hyperparameter analysis confirms the optimal parameters among various settings, and experiments on H\&M, and Netflix datasets indicate that the framework can adapt to differing recommendation domains. The experimental results confirm that integrating semantic clustering and uncertainty assessment yields more reliable and accurate recommendations.

\keywords{LLM for data management \and Uncertain data \and Information retrieval.}
\end{abstract}

\section{Introduction}

\begin{figure}[h]
    \centering
    \includegraphics[width=1\linewidth]{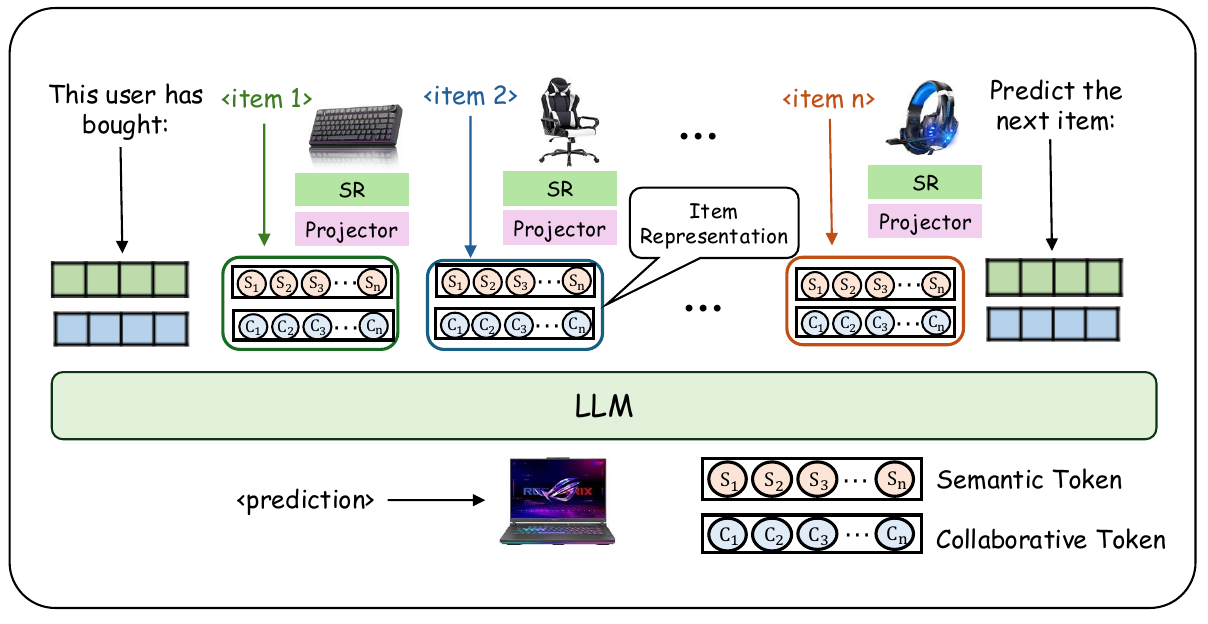}
    \caption{LLM-Based Sequential Recommendation Framework, which integrates sequential recommender (SR) and projector (e.g., a trainable MLP). These embeddings jointly form a collaborative token prompt (e.g., ``This user has bought: [item1], [item2], \ldots [item\(_n\)]. Predict the next item this user will buy.''). }
    \label{fig:llm rec}
\end{figure}

Large Language Models (LLMs) have proven valuable for recommendation systems by enhancing personalized suggestion generation across diverse domains \cite{geng2022recommendation,liTextAllYou2023}. Their strong natural language capabilities support efficient handling of multi-modal data like text descriptions, visual features, and categorical attributes. LLMs unify various information sources under one framework, enabling flexible, high-quality recommendations that adapt to complex user preferences \cite{geng2023vip5}. Figure~\ref{fig:llm rec} shows how these models \cite{liaoLLaRALargeLanguageRecommendation2024a} process user histories and item metadata in an autoregressive manner. However, current decoding strategies, inherited from natural language generation, can be inefficient for recommendation tasks that only require a single item identifier \cite{wangRethinkingLargeLanguage2024}. Beam search often entails unnecessary overhead, while greedy decoding may miss high-probability alternatives \cite{geng2023vip5}. This mismatch stems from framing item recommendations like text generation instead of recognizing unique structural requirements. Thus, specialized decoding methods are needed to address these domain-specific challenges.

This limitation leads us to examine recommendation tasks through the lens of semantic uncertainty, drawing inspiration from recent advances in natural language generation \cite{kuhn2023semantic} and recent work on uncertainty in recommender systems \cite{wang2023rethinkingmissingdataaleatoric,xiang2025exploit}. In sequential recommendation, we observe that user preferences often exhibit semantic equivalence patterns—different items or sequences can satisfy identical user needs despite having distinct representations. For instance, a user seeking wireless headphones might be equally satisfied with several similar products across different brands. This phenomenon parallels semantic equivalence in natural language, where different expressions can convey identical meanings. Standard decoding methods like beam search focus exclusively on token-level probabilities without considering these semantic relationships, potentially overestimating uncertainty when probability mass is distributed across semantically equivalent items. By explicitly modeling semantic uncertainty in user preferences, we can develop more effective decoding strategies that aggregate probabilities within semantic equivalence classes, leading to more accurate preference estimation and better recommendations.

Based on these insights, we propose an \textbf{U}ncertainty-aware \textbf{S}emantic \textbf{D}ecoding (\textbf{USD}) framework specifically designed for LLM-based sequential recommendation. Our framework consists of two key components: (1) A logit-based semantic clustering module that identifies semantically equivalent item groups by analyzing the similarity of their probability distributions in the user's preference space, significantly reducing redundant computations during decoding. (2) A semantic uncertainty estimation component that extends entropy calculation to capture uncertainty over preference patterns rather than individual tokens. By computing entropy over the meaning-distribution rather than token sequences, our approach provides a more accurate measure of recommendation uncertainty. The entire framework operates with minimal computational overhead, as semantic clustering is performed efficiently on a small set of candidate items, and uncertainty estimation is integrated seamlessly into the decoding process. Importantly, our method maintains compatibility with different LLM architectures and can be easily extended to handle multi-modal inputs through appropriate encoder designs.

\noindent The main contributions of this work are summarized as follows:
\begin{itemize}
    \item We identify fundamental limitations in conventional decoding strategies for LLM-based recommendation systems and introduce the Uncertainty-aware Semantic Decoding (USD) framework to address the misalignment between standard text generation approaches and recommendation requirements.
    
    \item We develop a logit-based semantic clustering algorithm that identifies preference equivalence patterns among candidate items and design an uncertainty-guided scoring mechanism that quantifies semantic entropy over clustered items rather than individual tokens.
    
    \item Extensive experiments on Amazon Product, H\&M, and Netflix datasets demonstrate that our method significantly outperforms state-of-the-art baselines, achieving improvements of 18.5\% in HR@3, 11.9\% in NDCG@3, and 10.8\% in MRR@3.
\end{itemize}

\section{Related Work}
\subsection{LLM for Recommendation}
Recent research shows that there is increasing interest in LLM4Rec. Several studies have examined the use of LLMs to predict user feedback through in-context learning~\cite{DBLP:conf/recsys/DaiSZYSXS0X23} and fine-tuning~\cite{TallRec}. These methods usually instruct LLMs to produce predictions directly without showing their intermediate reasoning steps, which limits the use of LLMs' reasoning ability. Additionally, recent work has used the reasoning capabilities of LLMs for recommendation tasks with prompting strategies such as chain-of-thought~\cite{CoT} and self-reflection~\cite{DRDT}. 
Despite these efforts, they often prioritize improving LLM performance through training enhancements, overlooking a crucial aspect: the detailed examination of the models’ recommendation outputs and the considerations for the decoding phase.

\subsection{LLM Uncertainty Quantification}
Uncertainty quantification (UQ) in Large Language Models (LLMs) is crucial for addressing model confidence and reliability~\cite{chen2023quantifying,fadeeva2024fact}. Token-level methods derive uncertainty from the probability distribution over generated tokens, often using entropy metrics to estimate response confidence~\cite{stengel2024lacie,he2020deberta}. Such approaches detect errors and improve interpretability. Another line of research, self-verbalized UQ, lets LLMs articulate their confidence, showing alignment with actual correctness~\cite{nikitin2024kernel,lieberum2024gemma}. Additionally, semantic similarity UQ checks multiple outputs for contradictions or hallucinations~\cite{dunefsky2024transcoders,ahdritz2024distinguishing,zhang2024target}, while mechanistic interpretability examines neuron-level decisions to illuminate model reasoning~\cite{chiang2023vicuna,wang2025knowledge,krause2023confidently,time25}. 

\subsection{LLM Decoding Generation}
LLMs employ varied decoding strategies to balance fluency, coherence, and factual soundness. A common approach is Maximum a Posteriori (MAP) decoding, selecting the most probable sequence~\cite{stahlberg-byrne-2019-nmt} but risking repetitive outputs. Self-refinement mechanisms iteratively revise generated text for improved coherence~\cite{madaan2023selfrefine,kumar2024traininglanguagemodelsselfcorrect}, while Minimum Bayes Risk (MBR) decoding focuses on minimizing expected risk~\cite{bertsch-etal-2023-mbr}. Recent work also explores compute-efficient solutions to lower inference cost~\cite{wu2024empiricalanalysiscomputeoptimalinference}. Another key line is guided decoding, relying on external knowledge to steer outputs toward factual correctness~\cite{hao2024llm,chen2023teaching}. Meta-search and ensemble-based methods further enhance diversity and accuracy by merging multiple generation paths~\cite{Havrilla2024GLoReWW,tyen2024llms}.

\section{Problem Formulation}
\noindent In this section, we establish the mathematical foundations for our semantic uncertainty-based decoding approach to LLM-powered sequential recommendation systems. \vspace{0.5em}

\noindent\textbf{Sequential Recommendation.} Let $\mathcal{U}$ denote the set of users and $\mathcal{I}$ the set of items. For each user $u \in \mathcal{U}$, their historical interaction sequence is denoted as $\mathcal{H}_u = \langle i_1, i_2, \ldots, i_n \rangle$, where each $i_j \in \mathcal{I}$ represents an interacted item. Given $\mathcal{H}_u$, the sequential recommendation task aims to predict the next item $i_{n+1}$ that the user is likely to interact with from $\mathcal{I}\setminus\mathcal{H}_u$, which represents the set of items the user has not yet interacted with.
\vspace{0.5em}

\noindent\textbf{LLM-based Decoding for Recommendation.} In LLM-based sequential recommendation, an item $i$ is represented through its contextual information $\mathcal{M}_i$ (e.g., title, category) and the prediction process uses an autoregressive formulation:
\begin{equation}
p_\theta(i_{n+1}|\mathcal{H}_u) = \prod_{t=1}^{T} p_\theta(y_t|y_{<t}, \mathcal{H}_u)
\end{equation}
\noindent where $p_\theta$ is the LLM with parameters $\theta$, and $y_t$ represents tokens in the output sequence. Traditional LLM approaches employ beam search decoding, which maintains multiple hypotheses throughout the generation process:
\begin{equation}
g(y|\mathcal{H}_u; p_\theta, \phi) \in \mathcal{P}(\mathcal{Y})
\end{equation}
\noindent where $g$ represents the generation algorithm with parameters $\phi$, and $\mathcal{P}(\mathcal{Y})$ is the space of probability distributions over possible outputs.
\vspace{0.5em}

\noindent\textbf{Semantic Equivalence in Item Space.} We formalize this through a semantic equivalence relation $E(s,s')$ between items:
\begin{equation}
E(s,s') = \mathbb{I}[\text{Sim}(s,s') > \tau]
\end{equation}
\noindent where $\text{Sim}(s,s')$ measures similarity between items based on their logit vectors, and $\tau$ is a threshold parameter. This relation partitions the item space into semantic clusters $C = \{c_1, c_2, ..., c_m\}$, where each cluster contains items that are functionally equivalent in satisfying specific user preferences.
\vspace{0.5em}

\section{Methodology}
\label{sec:method}
\begin{figure}[h]
    \centering
    \includegraphics[width=1\linewidth]{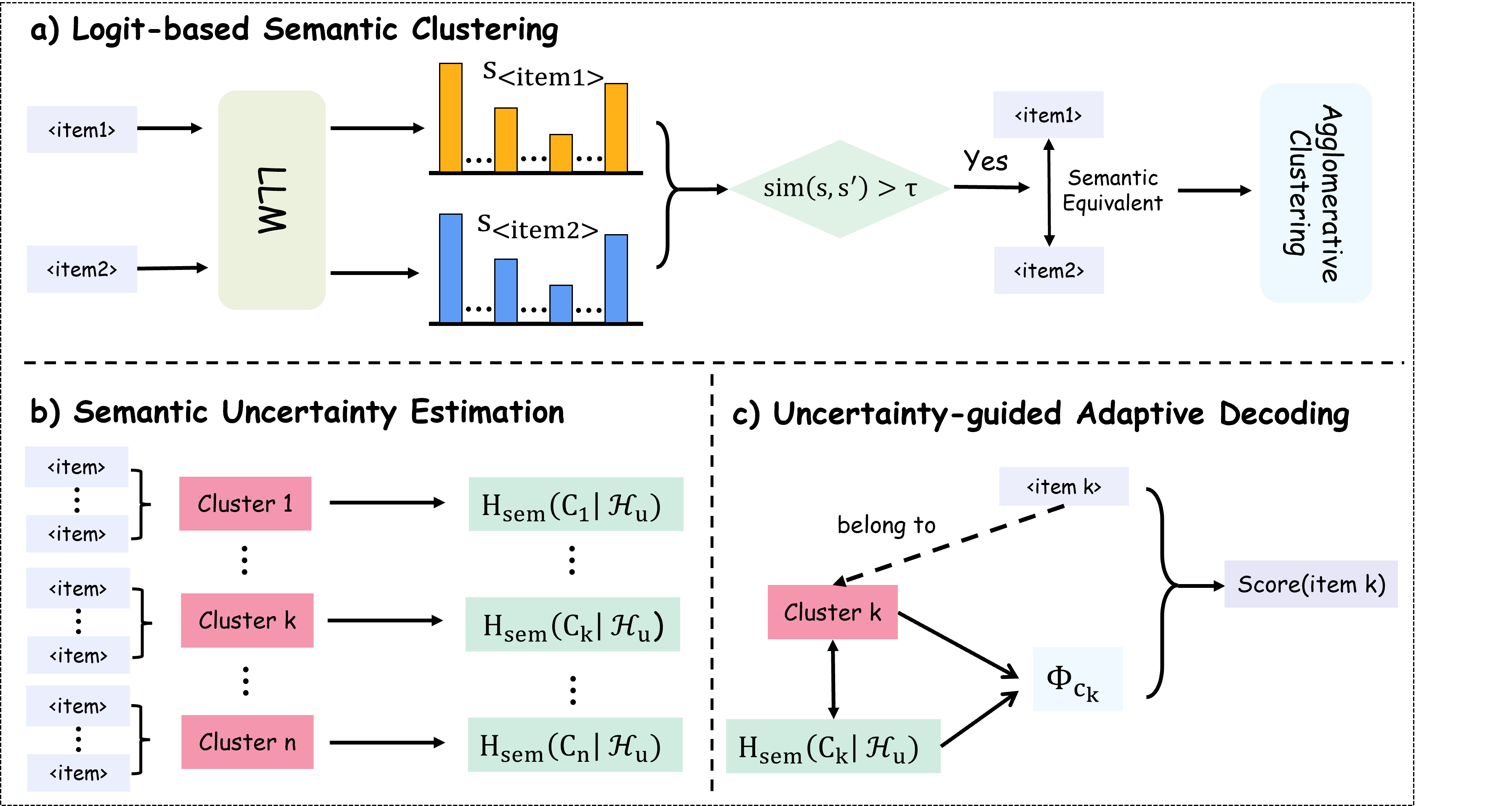}
    \caption{An overview of the proposed algorithmic framework for semantic clustering and uncertainty-guided decoding.}
    \label{fig:Algorithm framework}
\end{figure}

In this section, we introduce our Uncertainty-aware Semantic Decoding (USD) framework, whose aim is to bridge the gap between standard text-generation-oriented decoding methods and the unique needs of sequential recommendation. We begin with an overview of the main components and then detail each submodule: (i) \textbf{Logit-based Semantic Clustering} defines how candidate items form equivalence groups; (ii) \textbf{Semantic Uncertainty Estimation} quantifies user preference ambiguity by measuring entropy over these clusters; (iii) \textbf{Uncertainty-guided Adaptive Decoding} integrates semantic entropy information to balance exploration and exploitation. As depicted in Figure~\ref{fig:Algorithm framework}, our framework is designed to cluster candidate items based on their logit similarities, estimate uncertainty across these clusters, and then provide final item suggestions guided by an adaptive scoring mechanism.

\subsection{Logit-based Semantic Clustering}
Standard decoding processes seldom differentiate between closely related items that satisfy an identical user need. To alleviate such redundancy, we introduce a similarity measure:
\begin{equation}
\text{Sim}(s,s') \;=\; \frac{\mathbf{l}_s \cdot \mathbf{l}_{s'}}{\|\mathbf{l}_s\|\;\|\mathbf{l}_{s'}\|}.
\end{equation}
In this equation, $s$ and $s'$ are two candidate items drawn from the model's output space, $\mathbf{l}_s$ and $\mathbf{l}_{s'}$ denote their respective logit representations (i.e., pre-softmax activations) conditioned on the user history $\mathcal{H}_u$, and $\|\mathbf{l}_s\|$ represents the Euclidean norm of $\mathbf{l}_s$. The dot product $\mathbf{l}_s \cdot \mathbf{l}_{s'}$ therefore captures alignment in their logit-level embeddings. We then declare $s$ and $s'$ to be semantically equivalent if $\text{Sim}(s,s')$ exceeds a threshold:
\begin{equation}
E(s,s') \;=\; \mathbb{I}\bigl[\text{Sim}(s,s') \;>\; \tau\bigr],
\end{equation}
where $\tau \in [0,1]$ is the similarity threshold and $\mathbb{I}[\cdot]$ is the indicator function that returns 1 if the condition is true, 0 otherwise. We use an agglomerative procedure to cluster items satisfying $E(s,s')$, yielding a set of semantic clusters $C = \{c_1, c_2, \dots, c_m\}$ whose members fulfill comparable user intents.

\subsection{Semantic Uncertainty Estimation}
Based on these clusters, we measure preference ambiguity across clusters rather than individual items. Let $p_{\theta}(s \mid \mathcal{H}_u)$ be the LLM's probability of item $s$, given user history $\mathcal{H}_u$. We sum these probabilities over each cluster $c$:
\begin{equation}
p(c \mid \mathcal{H}_u) \;=\; \sum_{s \,\in\, c} p_{\theta}(s \,\mid\, \mathcal{H}_u).
\end{equation}
Here, $c$ represents a single cluster and $s$ enumerates the items residing within $c$, while $p_{\theta}(s \mid \mathcal{H}_u)$ comes directly from the original language model with parameters $\theta$. We then define the semantic entropy:
\begin{equation}
H_{\text{sem}}(C \,\mid\, \mathcal{H}_u) \;=\; - \sum_{c \,\in\, C} \bigl[p(c \mid \mathcal{H}_u)\,\log p(c \mid \mathcal{H}_u)\bigr].
\end{equation}
In this definition, $H_{\text{sem}}(C \mid \mathcal{H}_u)$ quantifies the overall uncertainty of user interests at the cluster level, $p(c \mid \mathcal{H}_u)$ is the probability assigned to each cluster, and the summation extends over all clusters $c \in C$. By concentrating probability mass on $c$, we avoid excessive dilution by semantically duplicative items. Monte Carlo sampling of $K$ candidate items helps compute $H_{\text{sem}}$ efficiently during inference.

\subsection{Uncertainty-guided Adaptive Decoding}
Where standard beam search or greedy strategies may misallocate computational effort, we propose a scoring function that incorporates semantic entropy. For each candidate item $s$ in the sampled set $S$, we define
\begin{equation}
\text{Score}(s) \;=\; (1 - \alpha)\,p_{\theta}(s \mid \mathcal{H}_u) + \alpha\,\Phi\bigl(s, C, H_{\text{sem}}\bigr).
\end{equation}
In this formula, $s$ is a single candidate item, $\alpha \in [0,1]$ is a hyperparameter that tunes the extent to which we trust item-level model probabilities versus cluster-level reasoning, and $p_{\theta}(s \mid \mathcal{H}_u)$ is the item's base probability. The function $\Phi(s, C, H_{\text{sem}})$ is defined as
\begin{equation}
\Phi(s, C, H_{\text{sem}}) \;=\; \frac{p(c_s \mid \mathcal{H}_u)}{|c_s|}\,\Bigl(1 - \beta\,H_{\text{sem}}(C\mid \mathcal{H}_u)\Bigr),
\end{equation}
where $c_s$ is the semantic cluster containing item $s$, $|c_s|$ denotes the size of that cluster, $\beta \in [0,1]$ specifies how much to factor in the entropy term, and $H_{\text{sem}}(C \mid \mathcal{H}_u)$ is the previously mentioned cluster-level entropy. Additionally, we adapt the sampling temperature:
\begin{equation}
\tau \;=\; \tau_0\bigl(1 + \gamma\,H_{\text{sem}}(C \mid \mathcal{H}_u)\bigr),
\end{equation}
where $\tau_0 > 0$ is a base temperature, $\gamma \ge 0$ regulates how strongly uncertainty drives exploration, and $H_{\text{sem}}$ is the same semantic entropy. In this manner, high uncertainty pushes the system to broaden candidate exploration, while low uncertainty narrows focus for precision.


\section{Experiments}

In this section, we evaluate our proposed Uncertainty-aware Semantic Decoding (USD) framework for sequential recommendation through comprehensive experiments across multiple real-world datasets. Our evaluation aims to answer the following research questions: \textbf{RQ1:} How does USD perform compared to state-of-the-art recommendation methods? \textbf{RQ2:} How much does each component contribute to USD's overall performance? \textbf{RQ3:} How does USD compare with alternative decoding strategies? \textbf{RQ4:} How well does USD generalize across different recommendation domains and datasets? \textbf{RQ5:} How do different hyperparameter settings affect USD's effectiveness?

\subsection{Experimental Setup}

\noindent\textbf{Datasets}
We evaluate our approach on multiple large-scale datasets from the Amazon Product Dataset\footnote{https://huggingface.co/datasets/McAuley-Lab/Amazon-Reviews-2023.}, spanning six diverse domains: Baby, Beauty, Clothing, Grocery, Sports, and Toys. For each product, we extract attributes including titles, descriptions, categories, and brand attributes. We filter out users and items with fewer than 10 interactions to ensure data quality. For evaluation, we adopt the leave-one-out strategy, using each user's last interaction for testing, second-to-last for validation, and the remaining interactions for training.


\noindent\textbf{Implementation Details}.We implement our USD framework using RedPajama-INCITE-Instruct-3B as the backbone model, which comprises 2.8B parameters with a decoder-only transformer architecture (32 layers). To optimize memory usage while maintaining numerical stability, we employ mixed precision training (FP16). For our USD framework hyperparameters, we set semantic similarity threshold $\tau=0.8$, uncertainty weight $\alpha=0.5$, and uncertainty parameter $\beta=0.3$ based on validation performance. For candidate generation, we use $K=10$ items with a temperature of $\tau_0=0.95$. We implement our model using PyTorch 2.0 and the Accelerate library for distributed evaluation on 8 NVIDIA RTX 3090 GPUs. 

\noindent\textbf{Baselines}. We compare against four categories of baselines, all baselines undergo identical data preprocessing and follow the same evaluation protocol:
\begin{itemize}
\item Non-LLM methods: Traditional collaborative filtering (MF \cite{koren2009matrix,rendle2012bpr}, LightGCN \cite{he2020lightgcn}), and sequential models (HGN \cite{ma2019hierarchical}, GRU4Rec \cite{jannach2017recurrent}, SASRec \cite{kang2018self}, S$^3$-Rec \cite{zhou2020s3}, MACR \cite{wei2021model}, UltraGCN \cite{mao2021ultragcn});
\item LLM-based approaches: Language models (BERT4Rec \cite{sun2019bert4rec}, UniSRec \cite{hou2022towards}, P5 \cite{geng2022recommendation}, VIP5$^+$ \cite{geng2023vip5}), and multi-modal variants (VBPR \cite{he2016vbpr}, CausalRec \cite{qiu2021causalrec}, MM-GCL \cite{yi2022multi}, MMSSL \cite{wei2023multi}, UniMP\cite{wei2024unimp}).
\end{itemize}

\subsection{Performance Comparison(RQ1)}
\label{sec:RQ1}

Our extensive evaluation demonstrates that USD effectively addresses the fundamental mismatch between standard LLM decoding strategies and recommendation requirements. 

\vspace{0.5em}
\begin{table*}[ht]
    \caption{Performance Comparison of Different Models}
    \label{tab:results}
    \centering

    \resizebox{0.8\textwidth}{!}{  
    \begin{tabular}{llcccccc}
        \toprule
        Type & Method & HR@3 & NDCG@3 & MRR@3 & HR@5 & NDCG@5 & MRR@5 \\
        \midrule
        \multirow{8}{*}[-2pt]{Non-LLM} 
            & MF & 0.0105 & 0.0077 & 0.0065 & 0.0165 & 0.0093 & 0.0078 \\
            & MACR & 0.0110 & 0.0080 & 0.0068 & 0.0170 & 0.0105 & 0.0091 \\
            & LightGCN & 0.0142 & 0.0103 & 0.0088 & 0.0206 & 0.0129 & 0.0094 \\
            & UltraGCN & 0.0151 & 0.0111 & 0.0095 & 0.0215 & 0.0134 & 0.0102 \\
            & HGN & 0.0167 & 0.0113 & 0.0113 & 0.0231 & 0.0153 & 0.0117 \\
            & GRU4Rec & 0.0132 & 0.0101 & 0.0086 & 0.0201 & 0.0128 & 0.0096 \\
            & SASRec & 0.0189 & 0.0124 & 0.0102 & 0.0276 & 0.0175 & 0.0126 \\
            & S$^3$-Rec & {0.0205} & 0.0149 & 0.0133 & 0.0311 & 0.0193 & 0.0156 \\
        \midrule
        \multirow{11}{*}[-5pt]{LLM-based}
            & BERT4Rec & 0.0121 & 0.0092 & 0.0079 & 0.0198 & 0.0127 & 0.0105 \\
            & UniSRec & 0.0201 & 0.0146 & 0.0135 & 0.0281 & 0.0191 & 0.0158 \\
            & P5 & 0.0086 & 0.0056 & 0.0042 & 0.0124 & 0.0074 & 0.0061 \\
            & VIP5$^+$ & 0.0175 & 0.0125 & 0.0108 & 0.0262 & 0.0163 & 0.0127 \\
            & VBPR & 0.0114 & 0.0084 & 0.0071 & 0.0181 & 0.0102 & 0.0086 \\
            & CausalRec & 0.0143 & 0.0105 & 0.0088 & 0.0229 & 0.0146 & 0.0121 \\
            & MMGCL & 0.0151 & 0.0112 & 0.0095 & 0.0241 & 0.0159 & 0.0130 \\
            & MMSSL & 0.0181 & 0.0132 & 0.0112 & 0.0281 & 0.0194 & 0.0164 \\
            & UniMP & 0.0248 & 0.0194 & 0.0176 & 0.0337 & 0.0231 & 0.0196 \\
            & USD (Ours) & \textbf{0.0294} & \textbf{0.0217} & \textbf{0.0195} & \textbf{0.0388} & \textbf{0.0257} & \textbf{0.0214} \\
        \bottomrule
    \end{tabular}}
\end{table*}
\vspace{0.5em}

Table~\ref{tab:results} presents comparisons against various state-of-the-art methods across six diverse domains from the Amazon Product Dataset. USD consistently outperforms all baselines, achieving average improvements of 18.5\% on HR@3, 11.9\% on NDCG@3, and 10.8\% on MRR@3 compared to the strongest baseline. Non-LLM methods like MF and LightGCN show limited performance due to their inability to leverage rich semantic information, while sequential models (SASRec, GRU4Rec) capture temporal patterns but fall short in modeling semantic relationships. Even advanced LLM-based methods (P5, UniMP) underperform USD as they rely on decoding strategies originally designed for text generation rather than recommendation tasks. These results confirm our motivation that standard decoding approaches fundamentally misalign with recommendation objectives. By explicitly modeling semantic relationships between items and incorporating uncertainty estimation, USD effectively bridges this gap, providing more accurate recommendations by recognizing when probability mass distributed across similar items represents a single strong preference rather than genuine uncertainty.

\subsection{Ablation Studies(RQ2)}
\label{sec:RQ2}
\vspace{0.5em}
\begin{table}[h]
\caption{Ablation study on key components of USD framework. }
\label{tab:ablation_components}
\centering
 \resizebox{0.8\textwidth}{!}{  

\begin{tabular}{lcccccccc}
\toprule
\textbf{Method Variant} & \textbf{HR@3} & \textbf{NDCG@3} & \textbf{MRR@3} & \textbf{HR@5} & \textbf{NDCG@5} & \textbf{MRR@5}  \\
\midrule
Complete USD & \textbf{0.0294} & \textbf{0.0217} & \textbf{0.0195} & \textbf{0.0388} & \textbf{0.0257} & \textbf{0.0214}  \\
USD w/o UE & 0.0273 & 0.0203 & 0.0182 & 0.0361 & 0.0240 & 0.0198  \\
USD w/o SC & 0.0263 & 0.0195 & 0.0176 & 0.0346 & 0.0230 & 0.0190  \\
\bottomrule
\end{tabular}}
\end{table}



Table~\ref{tab:ablation_components} reveals that removing any component significantly degrades performance, confirming their complementary nature. The semantic clustering (\textbf{SC}) component provides the most substantial contribution (10.5\% decrease in HR@3 when removed), followed by uncertainty estimation (7.1\% decrease). Removing semantic clustering causes the model to treat semantically similar items as distinct alternatives, leading to diluted probability estimates and suboptimal rankings. This confirms our core motivation that standard decoding methods fail to capture semantic relationships between items, resulting in probability mass being distributed across functionally equivalent items without appropriate aggregation. Similarly, disabling uncertainty estimation(\textbf{UE}) reduces the model's adaptability in ambiguous preference scenarios, particularly when users exhibit interest across multiple distinct categories. These findings validate our theoretical understanding of semantic equivalence in recommendation spaces and demonstrate the importance of explicitly modeling uncertainty in user preferences.

\begin{figure}[h]
    \centering
    \includegraphics[width=1\linewidth]{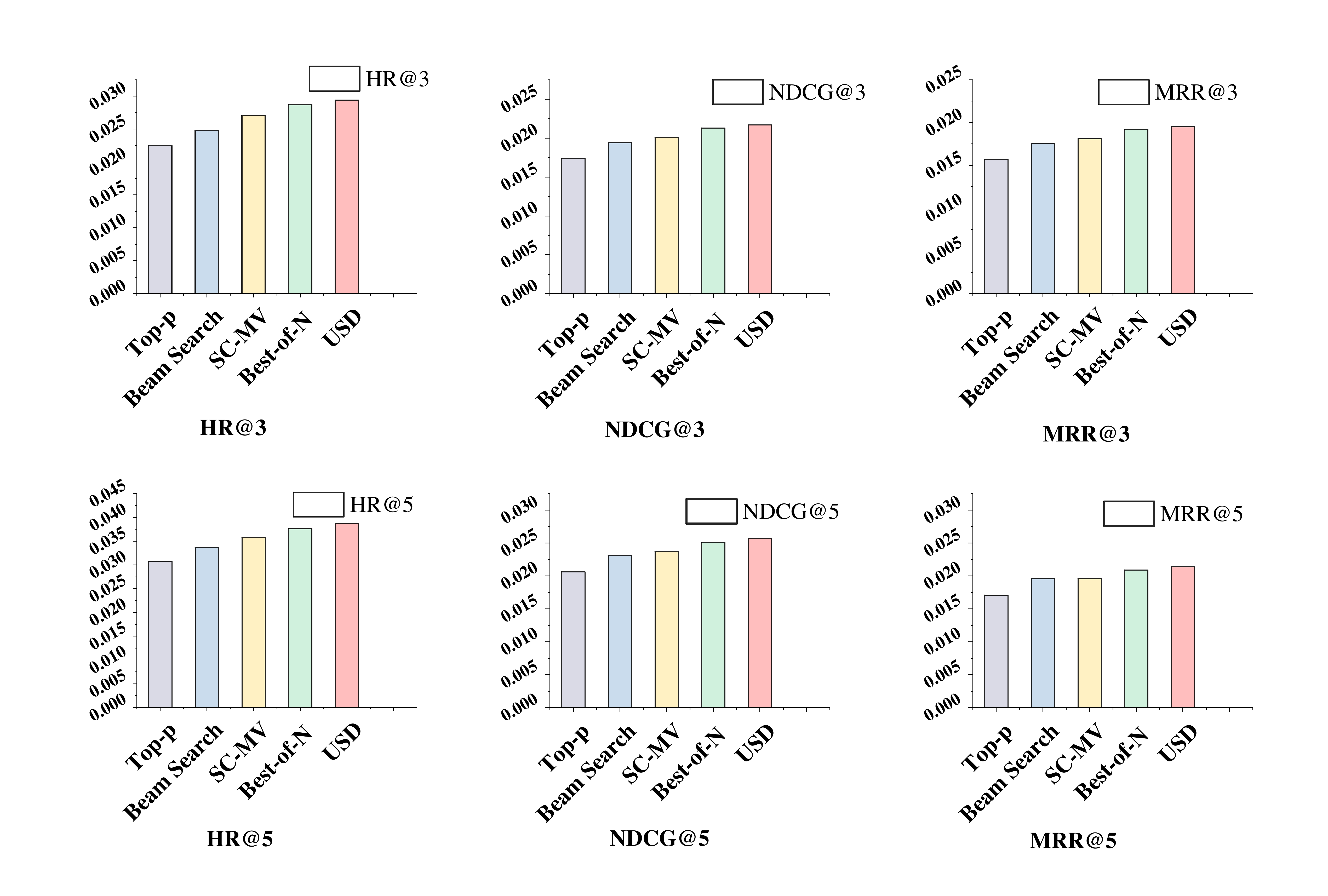}
    \caption{Decoding Strategies Comparison}
    \label{fig:decoding_comparison}
\end{figure}

\subsection{Decoding Strategies Comparison(RQ3)}
\label{sec:RQ3}

To demonstrate how USD addresses the mismatch between standard decoding strategies and recommendation requirements, we implement and compare four alternative decoding algorithms: Beam Search, Nucleus (Top-p) Sampling\cite{TOP-P}, Best-of-N\cite{bestofn2005,bestofn2009}. Self-consistency\cite{wang2023selfconsistency}. Figure~\ref{fig:decoding_comparison} shows that USD outperforms all alternatives across all metrics. Across multiple datasets, USD outperforms these baselines with HR@3 improvements ranging from 2.4\% to 23.5\%. Concretely, Beam Search (HR@3 around 0.0248) and Nucleus Sampling (0.0225) struggle to reassign probability among semantically similar items, while Best-of-N (0.0287) and Self-consistency (0.0271) improve coverage but lack explicit semantic modeling. By contrast, USD (0.0294) consolidates probabilities within logit-based clusters and leverages uncertainty-driven cues to more effectively produce the next recommendation.


\subsection{Cross-Domain Generalization Analysis(RQ4)}
\label{sec:RQ4}

Beyond our primary validation on the Amazon dataset, we examine USD's transferability to distinct recommendation domains: H\&M\footnote{https://www.kaggle.com/competitions/h-and-m-personalized-fashion-recommendations/data.} and Netflix\footnote{https://www.kaggle.com/datasets/netflix-inc/netflix-prize-data.} in Table~\ref{tab:hm and netflix}. These datasets present fundamentally different challenges—H\&M with its complex retail ecosystem and hierarchical product relationships, and Netflix with its content-centric preference patterns and temporal dynamics. 

\begin{table}[h]
  \caption{Performance Comparison on H\&M and Netflix Datasets}
  \label{tab:hm and netflix}
  \centering
  \resizebox{0.7\textwidth}{!}{  
    \begin{tabular}{lcccccc}
      \toprule
      & \multicolumn{3}{c}{H\&M} & \multicolumn{3}{c}{Netflix} \\
      \cmidrule(r){2-4} \cmidrule(l){5-7}
      & HR@5 & NDCG@5 & MRR@5 & HR@5 & NDCG@5 & MRR@5 \\
      \midrule
      P5 & 0.0101 & 0.0063 & 0.0046 & 0.0742 & 0.0413 & 0.0316 \\
      VIP5 & 0.0122 & 0.0118 & 0.0093 & 0.0936 & 0.0589 & 0.0395 \\
      S3-Rec & 0.0185 & 0.0121 & 0.0102 & 0.1155 & 0.0632 & 0.0511 \\
      UniSRec & 0.0196 & 0.0139 & 0.0107 & 0.1324 & 0.0856 & 0.0644 \\
      UniMP & 0.0313 & 0.0206 & 0.0172 & 0.1723 &  0.1196 & 0.1024 \\
      USD & \textbf{0.0368} & \textbf{0.0242} & \textbf{0.0203} & \textbf{0.1835} & \textbf{0.1271} & \textbf{0.1091} \\
      \bottomrule
    \end{tabular}
    }
\end{table}

The results reinforce our key findings from earlier experiments. USD consistently outperforms all baselines across both domains, achieving improvements of 17.6\% in HR@5, 17.5\% in NDCG@5, and 18.0\% in MRR@5 over UniMP on the H\&M dataset. For Netflix, USD shows improvements of 6.5\% in HR@5, 6.3\% in NDCG@5, and 6.5\% in MRR@5 over the same baseline. This cross-domain evaluation directly tests USD approach to addressing decoding policy mismatch can be generalized in the recommended environment.
\begin{figure}[h]
    \centering
    \includegraphics[width=1\linewidth]{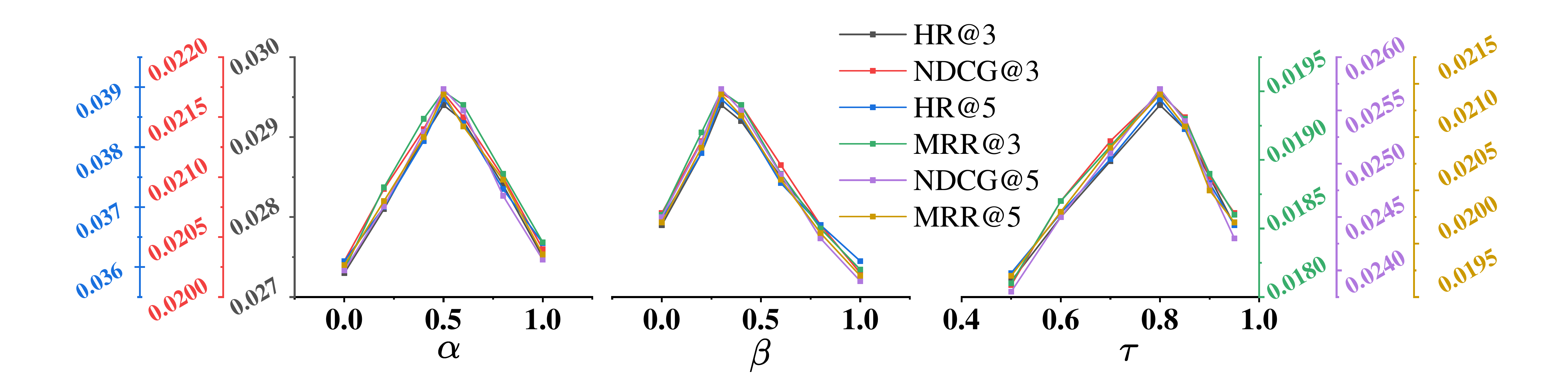}
    \caption{Hyperparameter Analysis}
    \label{fig:hypter}
\end{figure}

\subsection{Hyperparameter Analysis(RQ5)}
\label{sec:RQ5}

We conduct extensive hyperparameter sensitivity analyses to identify optimal configurations for USD across different domains in Figure~\ref{fig:hypter}. For the uncertainty weight parameter $\alpha$, which balances item-specific probability and semantic cluster information, performance peaks at $\alpha=0.5$ and declines symmetrically toward extremes. This pattern confirms our hypothesis that both individual item probabilities and semantic cluster information provide complementary signals for effective recommendation. When $\alpha=0$ (relying solely on item-specific probabilities), the model fails to capture semantic relationships between items—precisely the limitation of standard decoding approaches that motivated our work. Conversely, when $\alpha=1$ (relying solely on semantic clustering), the model loses precision by overlooking item-specific nuances. The uncertainty parameter $\beta$ achieves optimal results at $\beta=0.3$, representing an ideal balance between confidence in established preferences and exploration of alternatives under uncertainty. And the semantic similarity threshold $\tau=0.8$ creates clusters with optimal granularity. These parameter studies not only identify optimal configurations but also validate our core motivation: standard decoding approaches ($\alpha=0$) significantly underperform when they fail to account for semantic relationships, while our balanced approach ($\alpha=0.5$) effectively bridges the gap between token-level decoding and semantic-aware recommendation. 

\section{Conclusion}
In this work, we presented an Uncertainty-aware Semantic Decoding (USD) framework that addresses the core mismatch between traditional text-generation decoding and the specialized structure of sequential recommendation. By clustering semantically equivalent items at the logit level and quantifying uncertainty through semantic entropy, USD better allocates probability mass among near-duplicate candidates and refines exploration via an adaptive temperature mechanism. Extensive empirical evaluations reveal consistent improvements over state-of-the-art baselines, with gains exceeding 10\% on key metrics such as HR@3 and NDCG@3. Ablation analyses confirm that both semantic clustering and uncertainty estimation are crucial contributors to robust performance, while hyperparameter studies demonstrate the stability of the approach under varying threshold and entropy-weight settings. Future avenues may explore integrating domain-specific knowledge or multimodal signals, leveraging the framework’s capacity to seamlessly incorporate diverse contextual features for more personalized, efficient, and explainable recommendations.

\begin{credits}
\subsubsection{\ackname} 
This study was funded by the National Natural Science Foundation of China (Grant No. 72401232), the Natural Science Foundation of the Jiangsu Higher Educational Institution of China (Grant No. 23KJB520037), the XJTLU Research Development Fund (RDF-21-01-053), and Liaoning Liaohe Lab (Grant Nos. LLL24ZZ-02-01 and LLL24ZZ-02-02).
\end{credits}

\bibliographystyle{splncs04}
\bibliography{main}

\end{document}